# Absence of Tetragonal Distortion in $(1-x)$SrTiO$_3$-$x$Bi(Zn$_{1/2}$Ti$_{1/2}$)O$_3$ Solid Solution

Rishikesh Pandey, Ravi Kiran Pillutla, Uma Shankar and Akhilesh Kumar Singh

School of Materials Science and Technology

Indian Institute of Technology, Banaras Hindu University, Varanasi-221005, India

**Abstract**

We have carried out powder x-ray diffraction and dielectric studies on the lead free solid solution $(1-x)$SrTiO$_3$-$x$Bi(Zn$_{1/2}$Ti$_{1/2}$)O$_3$ [$(1-x)$ST-$x$BZT] with $x$=0.05, 0.10, 0.15, 0.20, 0.30 and 0.50 to explore the ferroelectric and piezoelectric properties. Analysis of the powder x-ray diffraction data reveals the cubic structure (space group Pm3m) of $(1-x)$ST-$x$BZT for the composition with $x \leq 0.20$, at room temperature as well as at low temperature. Highly tetragonal BZT fails to introduce any tetragonality when alloyed with ST. The solid solubility is limited to $x < 0.20$ and for higher BZT concentrations impurity phases start appearing. The ferroelectric and dielectric characterization of $(1-x)$ST-$x$BZT solid solution shows that all the compositions are paraelectric. The implications of the absence of tetragonal distortion in $(1-x)$ST-$x$BZT solid solution is discussed in connection with recently investigated other solid solutions based on BZT.

# I. INTRODUCTION

From the last several decades PbTiO$_3$ based morphotropic phase boundary solid solutions such as Pb(Zr$_x$Ti$_{1-x}$)O$_3$ (PZT), (1-x)Pb(Mg$_{1/3}$Nb$_{2/3}$)O$_3$-xPbTiO$_3$ (PMN-PT) etc. have been the materials of choice for electromechanical sensor, actuator and transducer applications, thanks to their excellent piezoelectric properties [1]. However, lead is toxic and hazardous for health and environment, and therefore, in recent years the quest for lead free piezoelectric ceramics with piezoelectric responses comparable to or better than PZT has been the prime focus of researchers [2]. To develop the lead free piezoceramics, some new solid solutions based on (Bi$_{1/2}$Na$_{1/2}$)TiO$_3$ [3], Ba(Zr$_{1-x}$Ti$_x$)O$_3$ [4] and K$_{1-x}$Na$_x$NbO$_3$ [5] have been explored but none of them exhibit the piezoelectric response that can match with the lead based counterparts. It is well known that the large piezoelectric response in PbTiO$_3$ (PT) based solid solutions is observed for the compositions close to the morphotropic phase boundary (MPB) which separates the tetragonal phase field of the PT rich compositions with the rhombohedral/monoclinic or pseudocubic compositions, the phase field for the component from the other side in the phase diagram [1]. Thus to develop a lead free solid solution which exhibit morphotropic phase boundary and high piezoelectric responses, a tetragonal non lead component is required which may be a promising substitute for PT so that one may expect a morphotropic phase boundary when it is alloyed with the components with rhombohedral or nontetrgonal structures [2]. In search of such a tetragonal substitute of PT, very recently a new Bi based compound Bi(Zn$_{1/2}$Ti$_{1/2}$)O$_3$ (BZT) has been investigated [6] with very high calculated ionic polarization ~103μC/cm$^2$ and large tetragonality (c/a~1.211) which is significantly greater than PT (c/a~1.06). It has been proposed that this compound has great potential to develop lead free morphotropic phase boundary systems with high piezoelectric response similar to Pb-based MPB systems [7-8]. The limitation with the BZT is that high pressure is required to synthesize the pure perovskite phase and formation of non perovskite phases start appearing when it is heated above 550 $^0$C at ambient pressure [6]. Thus BZT cannot be sintered and poled unless its solid solution is formed with some other component to reduce the large tetragonality. Recently several investigations have been carried out aiming to get MPB by solid solution formation of BZT with ferroelectric, BaTiO$_3$ (BT) [9], (Bi$_{1/2}$Na$_{1/2}$)TiO$_3$ (BNT) [10], (K$_{1/2}$Na$_{1/2}$)NbO$_3$ (KNN) [11], antiferroelectric PbZrO$_3$ (PZ) [12], and relaxor ferroelectric Pb(Mg$_{1/3}$Nb$_{2/3}$)O$_3$ (PMN) [13], etc. In contrast to PT based MPB systems, the addition of BZT in these components did not result in the morphotropic phase boundary. Further, the solid solution formation of BZT with PT increases the tetragonality of PT but, surprisingly solid solution formation of BZT with BT [9], BNT [10] and KNN [11] diminishes the long range ferroelectric order and leads to cubic structure with increasing BZT concentration. In view of this we decided to study the solid solution of BZT with $SrTiO_3$ (ST) which is well-known perovskite with cubic structure and is known to be very effective in decreasing tetragonality and transition temperature ($T_C$) of PT [14] and BT [15]. In the present work we have explored the (1-x)ST-xBZT solid solution with the objective to stabilize the perovskite phase of BZT and decrease its large tetragonality. However, we find that the cubic structure of ST persist in ST-BZT solid solution upto x=0.20 and then secondary non-perovskite phases start forming.

**II. EXPERIMENTAL DETAILS**

(1-x)ST-xBZT solid solution with compositions x=0.05, 0.10, 0.15, 0.20, 0.30 and 0.50 were prepared by conventional solid state ceramic method. Stoichiometric amounts of analytical reagent (AR) grade $Bi_2O_3$ (HIMEDIA, 99.5%), ZnO (QUALIGENS, 99%), $TiO_2$ (HIMEDIA, 99%), $SrCO_3$ (HIMEDIA, 99%) were used as raw materials. The powders were mixed by ball milling in acetone for 6 h. The mixed powder was dried and then calcined in a muffle furnace for 6 h at 900 $^0$C. The calcined powder was checked for phase purity by using 18 kW rotating anode Cu-target Rigaku (Japan) x-ray diffractometer operating in the Bragg Brentano geometry with the curved crystal graphite monochromator fitted in the diffracted beam. The instrumental resolution is 0.10 degree. The x-ray diffraction (XRD) data were collected at a scan rate of 2 degree/min in the 2- theta range from $20^0$- $120^0$ at the scan step of $0.02^0$. The calcined powder was mixed with binder (2% polyvinyl alcohol water solution) to form the pellets of thickness 1 to 1.5 mm using a cylindrical die (12 mm diameter) and uniaxial hydraulic press at an optimum load of 65 kN. These pellets were heated at 500 $^0$C for 10 h to burnout the binders. The pellets were sintered at 1100 $^0$C for 3h. Sintering of the samples was carried out in $Bi_2O_3$ atmosphere in a closed alumina crucible to avoid the $Bi_2O_3$ evaporation during sintering at high temperature. The $Bi_2O_3$ powder was used as spacer powder. Scanning electron microscopic examination was carried out by using ZEISS SUPRA40. Thin gold film was sputter coated on the sintered pellets before examining under SEM. For dielectric measurements, the flat surfaces of the sintered pellets were gently polished with 0.25μm diamond paste and then washed with acetone.

Isopropyl alcohol was then applied to clean the surfaces for removing the moisture, if any. Fired on silver paste was subsequently applied on both the surface of the pellet. It was first dried around 120 $^0$C in an oven and then cured by firing at 500 $^0$C for about 5 minutes. The frequency dependent dielectric constant ($\varepsilon^/$) and loss tangent (tan$\delta$) were measured at 1V signal strength by using Novocontrol (ALPHA A) high performance frequency analyzer. Radiant Ferroelectric Loop Tracer (USA) was used for P-E hysteresis measurement at 100Hz AC-field. The XRD data were analyzed by Rietveld profile refinement techniques using Fullprof package [16].

## III. RESULTS AND DISCUSSION

### A. Crystal Structure of (1-x)ST-xBZT

XRD patterns of (1-x)ST-xBZT ceramics with x=0.30 and 0.50 sintered at 1100 $^0$C is shown in the inset of Fig.1. We have marked the reflections corresponding to the perovskite structure by the letter 'P', in the XRD pattern. It is evident from this figure that the perovskite phase is formed for both these compositions. However, there are several peaks that could not be indexed by considering perovskite structure. A comparison of XRD patterns for x=0.30 and 0.50 indicates that several impurity peaks marked with asterisk, observed for x=0.50, are eliminated in the composition x=0.30. However, there are still few reflections that do not correspond to the perovskite phase. Due to difficulty in obtaining perovskite phase for the above two compositions (x= 0.50, 0.30), we decided to prepare compositions close to ST end. Fig.1 depicts the XRD patterns of sintered (1-x)ST-xBZT ceramics with x=0.05, 0.10, 0.15 and 0.20. For x=0.05, the diffraction pattern could be indexed by using cubic perovskite structure similar to that for ST. It suggests that for this composition, perovskite phase is stable. For compositions with x=0.10, 0.15, 0.20, most of the peaks correspond to perovskite phase but a few weak reflections marked with asterisk are also present which correspond to impurity phase. The secondary phase was identified as $Bi_2Ti_2O_7$ from the JCPDS file no. 32- 0118 for which the strongest peak appears at 2$\theta$= 29.9$^0$. The intensity of the reflections corresponding to this secondary phase increases with increasing 'x' i.e. concentration of BZT in the (1-x)ST-xBZT solid solution. The fraction of impurity phase estimated using intensity of the strongest peaks corresponding to the two phases is found to be very small (1.5% for x= 0.20). Thus, solid solubility of BZT in ST is very small (somewhere in between x=0.05 to 0.10) above which non-perovskite phase start forming. This is in marked contrast to the solid solution of ST with PT or BT where 100% solid solubility is observed. A close examination of XRD profiles shown in Fig.1 suggest that all the reflections are singlet, as there is no any distinguishable splitting of peaks is seen. Thus the structure of the (1-x)ST-xBZT solid solution with x ≤ 0.20 is also cubic similar to ST [14-15]. To confirm the cubic structure of different (1-x)ST-xBZT compositions prepared in the present work, we carried out full pattern Rietveld analysis of the XRD data with cubic structure in the Pm3m space group. As shown in Fig.2, very good fit between observed and calculated profiles is obtained for all the compositions of (1-x)ST-xBZT with x≤0.20. This confirms that the structure of (1-x)ST-xBZT is cubic in the Pm3m space group for the compositions with x≤ 0.20. The lattice parameter 'a' for x= 0.05, 0.10, 0.15 and 0.20 are 3.90516 Å, 3.90764 Å, 3.90932 Å and 3.90952 Å respectively. Thus there is very small increasing trend in 'a' with increasing BZT concentration in ST. The Ti-O bond length for 0.9ST-0.1BZT comes out to be 1.9528 Å which is comparable to the Ti-O2 bond length (1.976 Å) for tetragonal 0.9PT-0.1BZT but the Ti-O1 bond length (1.793 Å) is much different in the later as reported by Chen et al [17].

**B. Microstructure study**

Fig.3(a) and Fig.3(b) show the scanning electron microscope image of the pellet surface of the samples 0.95ST-0.05BZT and 0.70ST-0.30BZT, respectively, sintered at 1100 $^0$C. It is evident from Fig.3(a) that the microstructure is quite dense with the grains having cubical morphology and there is no signature of any impurity phase. This is in conformity with the results of the x-ray diffraction studies that reveal single phase cubic structure for 0.95ST-0.05BZT. Average grain size is found to be 0.6 $\mu$m. In contrast, for the composition 0.70ST-0.30BZT, two different types of morphology of the grains are clearly seen in Fig.3(b). The grains with flaky appearance are much bigger in size in comparison to smaller grains. This is consistent with the crystal structure studies which suggest presence of two phases for 0.70ST-0.30BZT. Looking at the microstructure shown in Fig.3(b) and also shown in Fig.3(a), we believe that the smaller size grains correspond to the perovskite phase while the grains with flaky appearance are due to impurity $Bi_2Ti_2O_7$ phase.

**C. Low temperature XRD studies**

As discussed previously, the structure of (1-x)ST-xBZT is cubic in the space group Pm3m for x≤ 0.20 at room temperature. Thus, no ferroelectric or piezoelectric behaviour is expected for these compositions at room temperature. It is reported that very small concentrations of BT or PT in ST can introduce ferroelectricity and tetragonality at low temperatures [14-15]. In view of this we decided to carry out low temperature XRD studies on 0.90ST-0.10BZT to check, if this transform to ferroelectric phase below room temperature? We find that the x-ray diffraction profiles of 0.90ST-0.10BZT remain singlet at 13 K also, similar to that at 300 K. This suggests that the structure of 0.1BZT- 0.90ST is cubic at 13 K also and there is no phase transition at low temperatures. Rietveld analysis of the powder XRD data for 0.90ST-0.10BZT also confirms the cubic structure at 13 K. Variation of lattice parameter with temperature for 0.90ST-0.10BZT is shown in Fig.4. As can be seen from the figure, except a slightly decreasing trend, there is no significant change in the lattice parameter with decreasing temperature from room temperature to down to 13 K. Thus, there is no significant thermal contraction or elongation in 0.90ST-0.10BZT at lower temperatures similar to the $(1-x)Bi(Mg_{1/2}Ti_{1/2})O_3-xPbTiO_3$ and iron nickel alloys [18, 19].

**D. Dielectric and Polarization Studies**

Fig.5 shows the frequency dependence of dielectric constant for (1-x)ST-xBZT composition with x=0.05, 0.10, 0.15 and 0.20. The room temperature dielectric constant for ST is 300. Addition of BZT in ST increases it up to ~ 480 for 0.95ST-0.05BZT even though the structure remains cubic. It further increases up to ~ 650 for the composition with x=0.10. However, it changes the nature and decreases significantly for the composition x=0.20 due to the appearance of impurity phases. As shown in Fig.5, the loss tangent is significantly small for each composition except 0.80ST-0.20BZT. The high value of tanδ at lower frequencies for 0.80ST-0.20BZT is due to the presence of impurity phase or barrier layer formation at the interface of the two phases.

We have also carried out polarization studies on these compositions to check if there is any signature of ferroelectric hysteresis loop. The P-E hysteresis loop for the composition with 0.20BZT- 0.80ST is shown in Fig.6. As can be seen from this figure, hysteresis loop is not saturated which confirms the absence of ferroelectric phase. Attempt to get the saturated hysteresis loop at higher field resulted into dielectric breakdown of the sample at the field strength of 60kV/cm. It is reported in thin films of Bi, Sr-based ferroelectric ceramics that significantly higher electric field (~200kV/cm) can be applied to get the hysteresis loop [20].

However, we do not expect a ferroelectric hysteresis loop in our samples even at higher fields as they possess centrosymmetric cubic crystal structure.

**E. Discussion**

The tetragonality of PT is 1.06 [21]. Substitution of ST in PT decreases the tetragonality and transition temperature but the solid solution exhibit tetragonal distortion for concentration of ST as high as 99.8% [14]. The tetragonality of BT is 1.02 [21]. Substitution of ST decreases the tetragonality and transition temperature but the tetragonal distortion of the crystal structure appears till 90% of ST [15]. However, in the case of BZT whose tetragonality is 1.211, significantly higher than BT and PT, there is no tetragonal distortion when solid solution is formed with ST. Absence of tetragonal distortion in (1-x)ST-xBZT ceramics is very much intriguing specially when we consider the fact that the substitution of BZT in PT increases the tetragonality of the solid solution [22]. Huang and Cann [23] have reported that solid solution formation of BZT with BT decreases the tetragonality and around 9% of BZT the structure becomes pseudocubic. In contrast to PT-BZT solid solution where $T_C$ increases with BZT [22], the $T_C$ of BT-BZT solid solution decreases with increasing BZT concentration and hysteresis loop becomes slim and dielectric peak gets diffused like relaxors [23]. In another report, on BT-BZT system, Wang and Yang [24] have shown that increasing the BZT concentration changes the symmetry from tetragonal to cubic for x=0.09. Dielectric and hysteresis loop characterization of the samples suggest that relaxor features are introduced in the BT-BZT solid solution with BZT substitution and for higher concentration the tetragonal distortion is lost [24]. A similar observation has been made by Wang et al also [25]. Wang et al [25] have further reported that substation of BZT in BNT diminishes the rhombohedral distortion resulting in cubic structure for even very less concentration of BZT (1%). Further, secondary non-perovskite phases start appearing for more than 10% of BZT concentrations. Similarly, the solid solution formation of BZT with KNN which is orthorhombic decreases the cell distortion and cubic symmetry appears for more than 30% BZT concentration [11]. The cubic to tetragonal transition temperature also decreases with increasing BZT concentration. Similarly, the $T_C$ decreases with increasing BZT concentration in PZ-BZT solid solution also [12]. Tailor et al [13] have reported that solid solution formation of BZT with the well known relaxor ferroelectric PMN (pseudocubic) does not introduce any tetragonal distortion, in marked contrast to PT, which decreases the relaxor features of PMN and transform it to normal ferroelectric with tetragonal structure for more than 35% PT [26]. In contrast, the highly tetragonal BZT increases the relaxor features of PMN as evidenced by increased frequency dispersion of the dielectric response [13] and the solid solution remains cubic. It should be pointed out here that though the increasing BZT concentration in PT increases the tetragonality and $T_C$ but the dielectric peaks becomes diffuse indicating appearance of relaxor features [22]. It is difficult to understand why the BZT, which increases the tetragonality of PT by solid solution formation, lowers the tetragonal/ferroelectric distortion of the crystal structure when alloyed with the other compounds like BT [9], BNT [10], KNN [11], antiferroelectric PZ [12], unless we consider the local disorder for BZT and solid solutions based on it. Enhancement of tetragonality for PT-BZT may be related with the presence of similar type of lone pair of electrons for $Pb^{2+}$ and $Bi^{3+}$ ions at A-site while for the other solid solution systems the situation is different. As reported by Suchomel et al [6], the large tetragonal distortion of BZT requires unusual coordination environment at both the A and B-sites that makes the B-site coordination geometry from tetragonally elongated octahedral to square based pyramidal. Further, the Rietveld structure refinement of neutron powder diffraction data by Chen et al [17] have shown that in 0.7PT-0.3BZT, the Zn exhibit more polarizable property than Ti that results in splitting of B-site cations by 0.27Å along the c-axis. This is also corroborated by the first principle calculations on BZT by Qi et al [27] that suggests high degree of local disorder for A and B-site cations. The density functional theory (DFT) calculations on BZT and PT-BZT have also shown that the Zn displacement is always different than the Ti and can be significantly larger than it [7, 8]. The large difference between Zn and Ti ionic charges, and also the displacements, may lead to generation of random fields that can prevent development of long range ferroelectric order and give rise to pseudo cubic structure [28]. Alloying BZT with PT which is strongly normal ferroelectric and also the fact that $Bi^{3+}$ and $Pb^{2+}$ have similar lone pairs of electrons might be able to accommodate the large local disorder so that the tetragonality of the solid solution increases with increasing BZT. However, increasing the BZT concentration the peak in the temperature dependence of the dielectric constant becomes very diffused similar to relaxors. The solid solutions based on BZT with other compounds may not be able to overcome the large local disorder of BZT and also the disorder being introduced by different valance and size of the other ions at A- and B-sites. This can prevent the long range ferroelectric distortion and results in pseudo-cubic structure with increasing BZT concentration. This can also lead to enhancement of relaxor characteristics as reported for several BZT based solid solutions with BT, BNT, KNN etc. It is well known that local and compositional disorder is responsible for the appearance of relaxor features and pseudocubic crystal structure [28]. In view of the foregoing discussion, we may conclude that BZT is not a good candidate material for developing lead free MPB systems with high piezoelectric response. As such there is no direct evidence of ferroelectricity in BZT and possibly it can be considered a pyroelectric for which there is no polarization switching by external electric field [6].

## IV. SUMMARY

Structural analysis of various compositions of (1-x)ST-xBZT ceramics reveals that the structure is cubic for the compositions with x≤0.20 in the Pm3m space group. The solid solubility limit of BZT in ST is less than 10% beyond which secondary phases start appearing. The cubic structure of the solid solution does not transform into any ferroelectric phase at low temperatures. We hope that this work will further stimulate the theoretical studies to understand the strange behaviour of BZT as to why it increases the $T_C$ when alloyed with PT while decreases the $T_C$ when solid solutions are formed with BT, BNT, KNN, etc.

## ACKNOWLEDGEMENT

R. P. acknowledges University Grant Commission (UGC), India for financial support as senior research fellowship (SRF).

REFERENCES:

1. S. E. Park and T. R. Shrout, J. Appl. Phys. **82**, 1804 (1997).
2. J. Rodel, W. Jo, K. T. P. Seifert, E. M. Anton, T. Granzow and D. Damjanovic, J. Am. Ceram. Soc. **92**, 1153 (2009).
3. W. C. Lee, C. Y. Huang, L. K. Tsao and Y. C. Wu, J. Eur. Ceram. Soc. **29**, 1443 (2009).
4. T. Maiti, R. Guo and A. S. Bhalla, J. Am. Ceram. Soc. **91**, 1769 (2008).
5. Y. Dai, X. W. Zhang and K. P. Chen, Appl. Phys. Lett. **94**, 042905 (2009).
6. M. R. Suchomel, A. M. Fogg, M. Allix, H. Niu, J. B. Claridge and M. J. Rosseinsky, Chem. Mater. **18**, 4987 (2006).
7. I. Grinberg, M. R. Suchomel, W. Dmowski, S.E. Mason, H. Wu, P. K. Davies and A. M. Rappe, Phys. Rev. Lett. **98**, 107601 (2007).

8. H. Wang, H. Huang, W. Lu, H. L. W. Chan, B. Wang and C. H. Woo, J. Appl. Phys. **105**, 053713 (2009).
9. C. C. Huang, D. P. Cann, X. Tan and N. Vittayakorn, J. Appl. Phys. **102**, 044103 (2007).
10. S. T. Zhang, F. Yan and B. Yang, J. Appl. Phys. **107,** 114110 (2010).
11. M. Sutapun, C. C. Huang, D. P. Cann and N. Vittayakorn, J. Alloy Compd. **479**, 462 (2009).
12. O. Khamman, X. Tan, S. Ananta and R. Yimnirun, J. Mater. Sci. **44**, 4321 (2009).
13. H. N. Tailor, A. A. Bokov and Z. G. Ye, Ferroelectrics **405**, 67 (2010).
14. V. V. Lemnov, E. P. Smirnova and E. A. Tarakanov, Phys. Solid. Stat. **39**, 628 (1997), F. Zhang, T. Karaki and M. Adachi, Powder Tech. **159**, 13 (2005).
15. C. Menoret, J. M. Kiat, B. Dkhil, M. Dunlop, H. Dammak and O. Hernandez, Phys. Rev. B **65**, 224104 (2002).
16. J. Rodriguez-Carvajal, FULLPROF, A Rietveld Refinement and Pattern Matching Analysis Program, Laboratory Leon Brillouin (CEA-CNRS), France, 2011.
17. J. Chen, X. Sun, J. Deng, Y. Zu, Y. Liu, J. Li and X. Xing, J. Appl. Phys. **105**, 044105, (2009).
18. P. Hu, J. Chen, X. Sun, J. Deng, X. Chen, R. Yu, L. Qiao and X. Xing, J. Mater. Chem. **19**, 1648 (2009).
19. M. Schilfgaarde, I. A. Abrikosov and B. Johansson, Nature **400**, 46 (1999).
20. S. Bhattacharyya, A. Laha and S. B. Krupanidhi, J. Appl. Phys. **91**, 4543 (2002); S. T. Zhang, B. Yang, X. J. Zhang, Y. F. Chen, Z. G. Liu and N. B. Ming, Mater. Lett. **56**, 221 (2002).
21. B. Jaffe, W. R. Cook and H. Jaffe, Piezoelectric Ceramics, Academic press, London (1971).
22. M. R. Suchomel and P. K. Davies, Appl. Phys. Lett. **86**, 262905 (2005).
23. C. C. Huang and D. P. Cann, J. Appl. Phys. **104**, 024117 (2008).
24. X. Wang and A. Yang, J. Phys. D: Appl. Phys. **42**, 075419 (2009).
25. L.Wang, J. H. Chao, Y. S. Sung, M. H. Kim, T. K. Song, S. S. Kim and B. C. Choi, Ferroelectrics **380**, 177 (2009).
26. A. K. Singh, D. Pandey and O. Zaharko, Phys. Rev. B **74**, 024101 (2006), A. K. Singh and D. Pandey, Phys. Rev. B **67**, 064102 (2003).

27. T. Qi, I. Grinberg and A. M. Rappe, Phys. Rev. B, **79**, 094114 (2009).

28. A. A. Bokov and Z. -G. Ye, J. Mater. Sci. **41**, 31 (2006).

**FIGURE CAPTIONS:**

FIG.1. Powder XRD patterns of (1-x)ST-xBZT ceramics with x=0.05, 0.10, 0.15, 0.20, 0.30 and 0.50 sintered at 1100 $^0$C for 3h. The inset shows the XRD patterns for x=0.30 and 0.50. The Miller indices given on top of the reflections correspond to cubic perovskite structure while the peaks marked with asterisk correspond to impurity $Bi_2Ti_2O_7$ phase.

FIG.2. Observed (dots), calculated (continuous line) and difference (continuous bottom line) profiles for (1-x)ST-xBZT ceramics with x= 0.05, 0.10, 0.15 and 0.20 obtained after Rietveld analysis of the powder XRD data using cubic space group Pm3m. The vertical tick marks above the difference plot show the peak positions. The peaks marked with asterisk correspond to impurity $Bi_2Ti_2O_7$ phase.

FIG.3. SEM image of the sintered pellet surface for (a) 0.95ST-0.05BZT and (b) 0.70ST-0.30BZT ceramic. The smaller size grains in (b) correspond to the perovskite phase while the bigger grains with flaky appearance are due to the impurity $Bi_2Ti_2O_7$ phase.

FIG.4. Variation of lattice parameter with temperature for 0.90ST-0.10BZT ceramics.

FIG.5. The frequency dependence of room temperature dielectric Permittivity and tanδ for (1-x)ST-xBZT ceramics with x=0.05, 0.10, 0.15 and 0.20.

FIG.6. P-E hysteresis loop for 0.20BZT- 0.80ST ceramic.

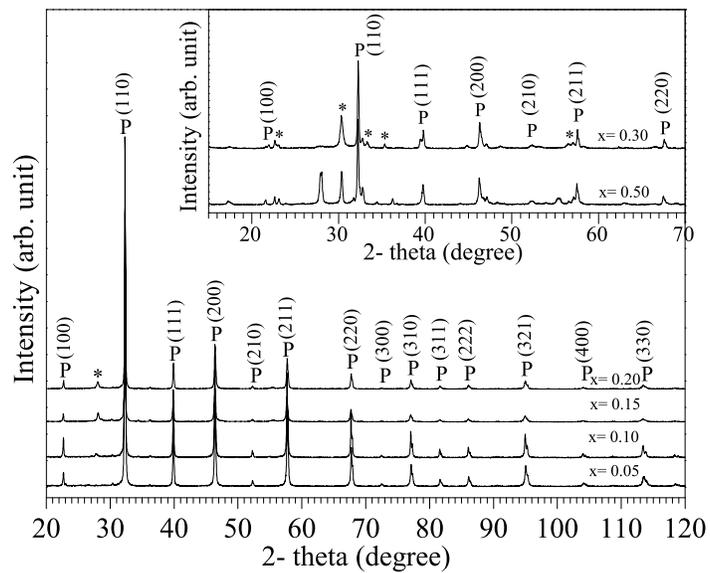

Fig.1

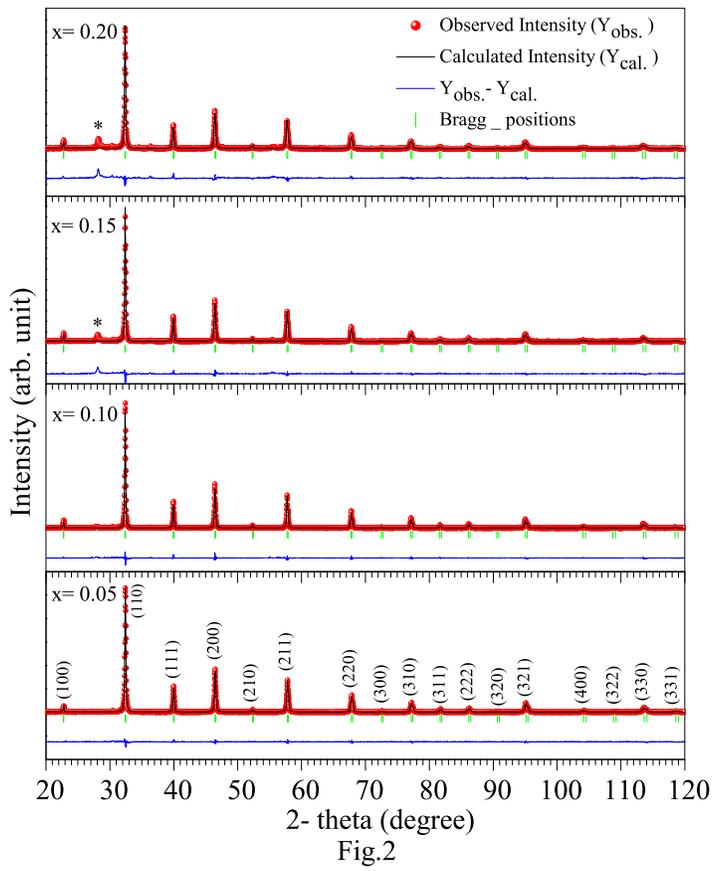

Fig.2

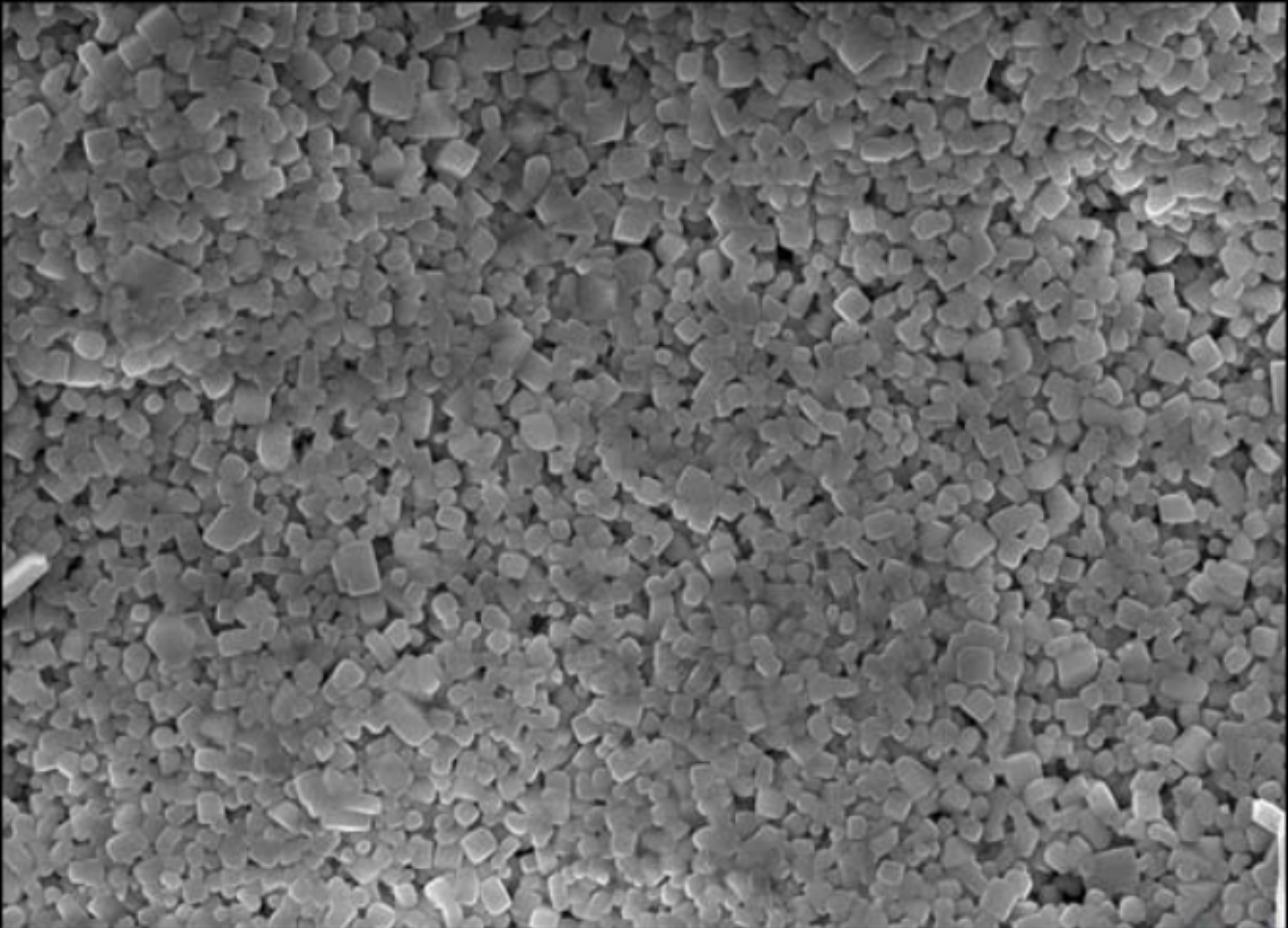

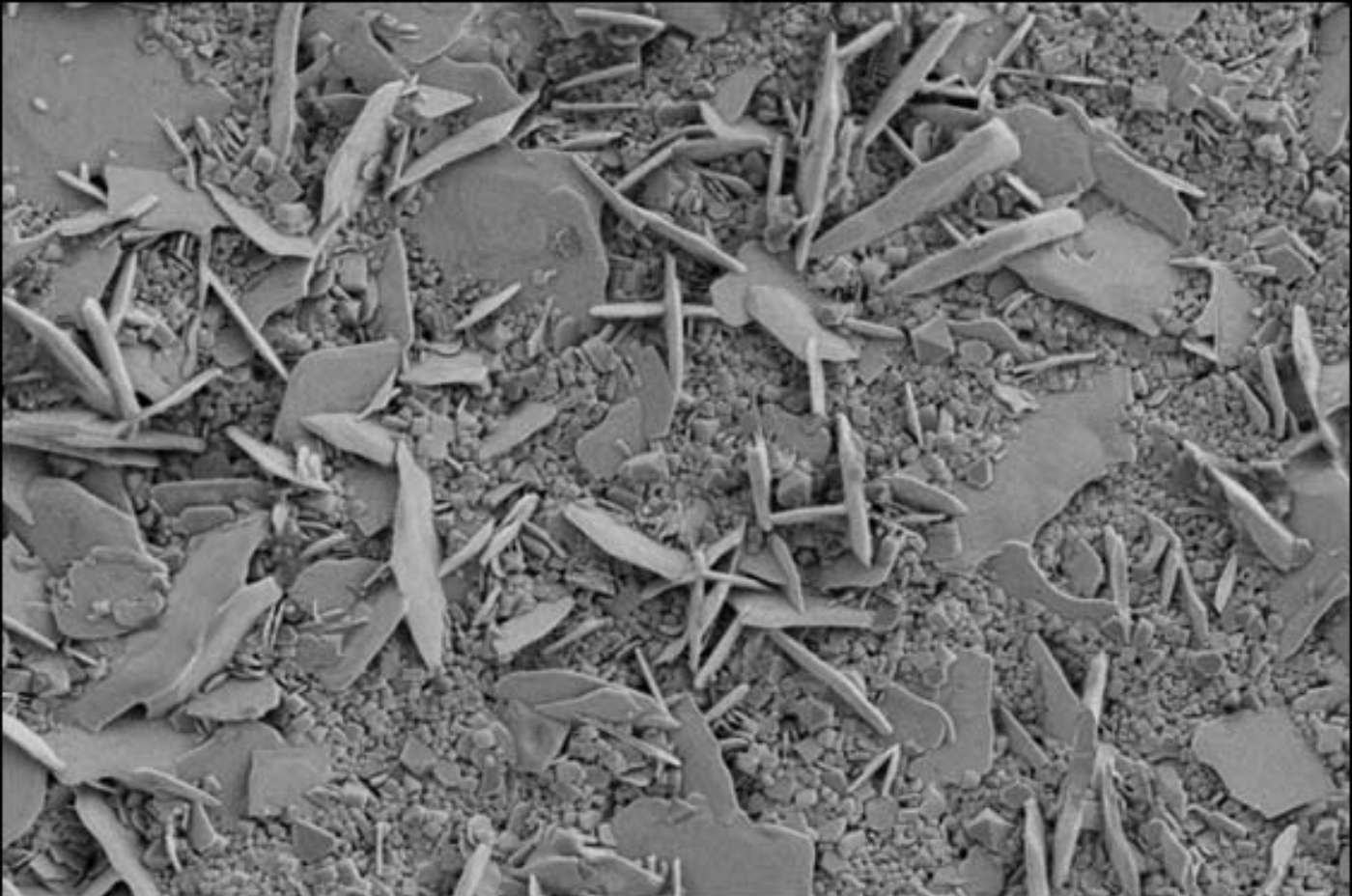

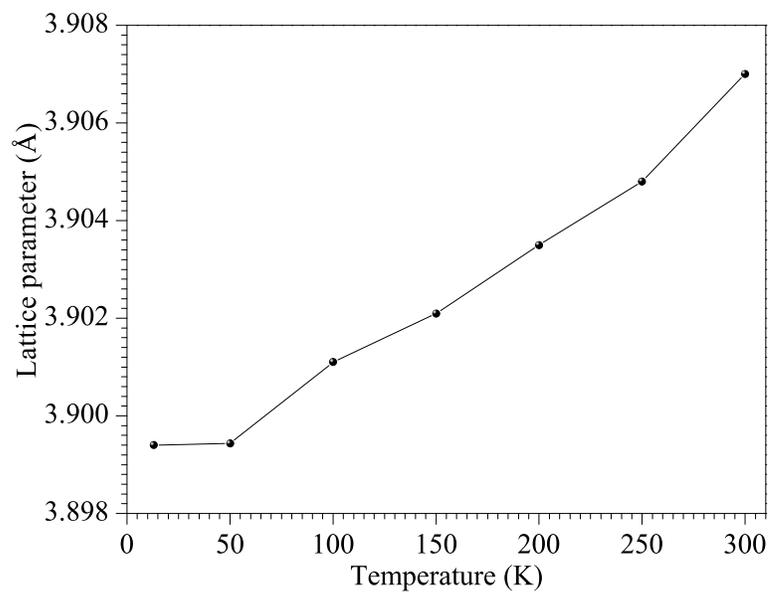
Fig.4

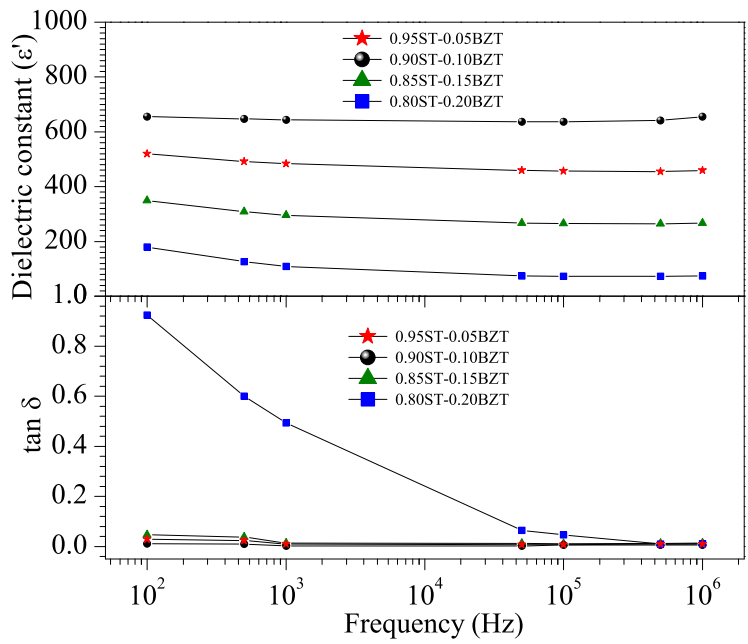

Fig.5

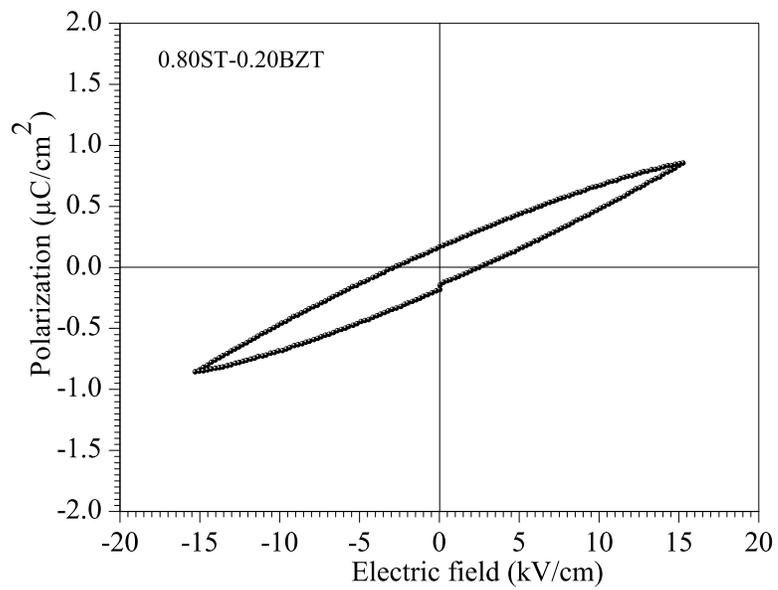

Fig.6